\documentclass[fleqn,usenatbib]{mnras}

\usepackage{newtxtext,newtxmath}

\usepackage[T1]{fontenc}
\usepackage{ae,aecompl}

\usepackage{graphicx}	
\usepackage{amsmath}	
\usepackage{amssymb}	

\usepackage{amssymb,xcolor}
\usepackage{colortbl}

\newcommand{\swift}{{\it Swift}}

\newcommand{\integral}{{\it INTEGRAL}}
\newcommand{\fermi}{{\it Fermi}}
\newcommand{\grs}{GRS 1716--249}
\newcommand{\gro}{GRO J1719--24}
\newcommand{\Cyg}{Cyg X--1}

\newcommand{\gx}{GX 339--4}
\newcommand{\maxi}{MAXI J1836--194}

\newcommand{\cm}{cm$^{-2}$}
\newcommand{\ergs}{erg\,s$^{-1}$}

\newcommand{\ergcm}{erg\,cm$^{-2}$\,s$^{-1}$}

\newcommand{\kte}{kT$_{\rm e}$}

\newcommand{\rms}{\textit{rms}}
\newcommand{\ls}{$l_{s}$}
\newcommand{\lh}{$l_{h}$}
\newcommand{\lnth}{$l_{nth}$}
\newcommand{\lth}{$l_{th}$}
\newcommand{\lb}{$l_{B}$}
\newcommand{\lc}{$l_{c}$}
\newcommand{\los}{$l_{obs}$}
\newcommand{\pj}{$P_{\rm jet}$}

\title[The soft $\gamma$-ray emission in \grs]{On the nature of the soft $\gamma$-ray emission in the hard state of the black hole transient GRS 1716--249}

\author[T. Bassi et al.]{
T. Bassi$^{1,2,3}$\thanks{E-mail: tiziana.bassi@inaf.it},
J. Malzac$^{3}$,
M. Del Santo$^{1}$,
E. Jourdain$^{3}$,
J.-P. Roques$^{3}$,
A. D'A\`i$^{1}$, 
\newauthor
J.C.A. Miller-Jones$^{4}$,
R. Belmont$^{5}$,
S.E. Motta$^{6}$, 
A. Segreto$^{1}$,
V. Testa$^{7}$,
P. Casella$^{7}$
\\
$^{1}$ INAF -- Istituto di Astrofisica Spaziale e Fisica Cosmica di Palermo, Via Ugo La Malfa 153, I-90146 Palermo, Italy\\
$^{2}$ Universit\'a degli Studi di Palermo -- Emilio Segr\'e, Dipartimento di Fisica e Chimica, via Archirafi 36,  I-90123 Palermo, Italy\\
$^{3}$ IRAP, Universit\'{e} de Toulouse, CNRS, UPS, CNES, Toulouse, France\\
$^{4}$ International Centre for Radio Astronomy Research -- Curtin University, GPO Box U1987, Perth, WA 6845, Australia\\
$^{5}$ AIM, CEA, CNRS, Universit\'e Paris-Saclay, Universit\'e Paris Diderot, Sorbonne Paris Cit\'e, 91191 Gif-sur-Yvette, France\\
$^{6}$ Department of Physics, Astrophysics, University of Oxford, Denys Wilkinson Building, Keble Road,\\ 
 OX1 3RH Oxford, UK 0000-0002-6154-5843\\
$^{7}$ INAF, Osservatorio Astronomico di Roma, Via Frascati 33, 00078 Monteporzio Catone, Roma, Italy\\
}

\date{Accepted XXX. Received YYY; in original form ZZZ}

\pubyear{}

\begin{document}
\label{firstpage}
\pagerange{\pageref{firstpage}--\pageref{lastpage}}
\maketitle

\begin{abstract}
The black hole transient \grs~was monitored from the radio to the $\gamma$-ray band during its 2016--2017 outburst. This paper focuses on the Spectral Energy Distribution (SED) obtained in 2017 February--March, when \grs~was in a bright hard spectral state.
The soft $\gamma$-ray data collected with the \integral/SPI telescope show the presence of a spectral component which is in excess of the thermal Comptonisation emission.
This component is usually interpreted as inverse Compton emission from a tiny fraction of non-thermal electrons in the X-ray corona. 
We find that hybrid thermal/non-thermal Comptonisation models provide a good fit to the X/$\gamma$-ray spectrum of \grs. The best-fit parameters are typical of the bright hard state spectra observed in other black hole X-ray binaries.
Moreover, the magnetised hybrid Comptonisation model {\sc belm} provides an upper limit on the intensity of the coronal magnetic field of about 10$^6$\,G.
Alternatively, this soft $\gamma$-ray emission could originate from synchrotron emission in the radio jet. In order to test this hypothesis, we fit the SED with the irradiated disc plus Comptonisation model combined with the jet internal shock emission model {\sc ishem}.  
We found that a jet with an electron distribution of $p\simeq2.1$ can reproduce the soft $\gamma$-ray emission of \grs.
However, if we introduce  the expected  cooling break around 10\,keV, the jet model can no longer explain the observed soft $\gamma$-ray emission, unless the index of the electron energy distribution is significantly harder ($p<2$).
\end{abstract}




\section{Introduction}
Black hole X-ray binaries (BHBs) are binary systems where a black hole accretes material from  an ordinary star. In these  systems the X-ray emission originating from the accretion flow can reach X-ray luminosities of L$_{\rm X}\sim 10^{36-39}$\,\ergs. While a few BHBs are persistent (characterised by steady X-ray emission), most of them are transient sources (called black hole transients, BHTs), i.e.\ they alternate long periods in a low-luminosity quiescent state with episodic outbursts. Furthermore, BHBs are associated with powerful jets whose synchrotron emission is usually observed in the radio and infrared bands.  
Based on X-ray spectral and timing properties, it is possible to identify characteristic spectral states\\ \citep{zdz04,remillard06}. In particular, the hard state (HS) is characterised by a dominating hard-X energy spectrum, interpreted as thermal Comptonisation of soft disc photons by a hot plasma (kT$_{\rm e}\sim$50--100\,keV) located close to the BH \citep{zdz04}.
A weak soft thermal component (inner disc black-body temperature of $\sim$0.1--0.2\,keV), possibly originating from the accretion disc truncated at large radii (roughly 100\,R$_{\rm g}$) from the BH \citep{done07}, is also observed.
A high fractional root mean squared (\rms) variability \citep[as high as 30\%,][]{munoz11, belloni16} also characterises hard spectral states.  

A flat or slightly-inverted  (F$_{\nu} \propto$ $\nu^{\alpha}$ with $\alpha \sim$0.5--0) radio/infrared (IR) spectrum is generally observed in the HS. This is interpreted as partially self-absorbed synchrotron emission from relativistic electrons ejected from the system through collimated compact jets \citep{corbel00,corbel02,fender00}, analogous to what is observed in active galactic nuclei \citep[AGN,][]{blandford79}. 
However, a dissipation mechanism compensating for the adiabatic losses is required to maintain the flat spectral shape, otherwise we would observe a strongly inverted radio spectrum.
\cite{malzac13,malzac14} showed that the internal shocks caused by fast fluctuations of the jet velocity can be an effective dissipation mechanism along the jets. The origin of the fluctuations is likely  driven by the variability of the accretion flow. In this framework, \cite{drappeau17} suggested that the drop of the radio/IR emission in the soft spectral state, usually ascribable to the quenching of the jets \citep{fender99, corbel00}, might be associated with the low X-ray variability (\rms~< 5\%) that characterises this spectral state.  

In recent years, a further component has been observed by the \integral~satellite in a number of BHBs, in addition to the thermal Comptonisation emission dominating the X-ray spectra of the hard states  \citep[e.g.][]{delsanto08, bouchet09, delsanto16}.
The first evidence of soft $\gamma$-ray emission above $\sim$200\,keV was observed in a few bright BHBs \citep{nolan81,johnson93,roques94}. Moreover, observations performed with the COMPTEL telescope showed that this component extends above 1\,MeV in \Cyg~\citep{mcconnell94,mcconnell02}.
Usually, this component is explained as a Comptonisation process due to a non-thermal electron population in the corona \citep{poutanen98,coppi99,gierlinski99,malzac09}. 
Alternative scenarios invoke proton-proton $\pi^{0}$ decay emission \citep{jourdain94}, or spatial/temporal variation of the thermal electrons' plasma parameters \citep{malzac00}.
Recent measurements of strongly-polarised emission above 250\,keV in \Cyg~have provided support to the hypothesis that a synchrotron process in the jet environment is a possible origin of this component \citep{laurent11, jourdain12, rodriguez15}. 
Studies performed on \Cyg~in the \fermi~energy band (above a few hundred MeV) have shown that this emission can be explained in terms of synchrotron self-Compton (SSC) processes \citep{zdz17}. 
On the other hand, it can also be well reproduced by jet models assuming continuous particle acceleration along the jet, even though a very hard electron energy distribution (electron index $\sim1.5$) and strong constraints on the magnetic field are required \citep{zdz14}.

The black hole transient \grs~ (\gro~ or Nova Oph 1993) was discovered by the {\it CGRO}/BATSE and {\it Granat}/SIGMA telescopes during an outburst that occurred in 1993 September \citep{harmon93,ballet93}. 
The orbital period was estimated as 14.7\,hr and a lower limit of 4.9\,M$_{\odot}$ for the mass of the compact object was reported by \citet{masetti96}, confirming the BHT nature of the source.  
Radio observations indicated a flat radio spectrum \citep{dellaValle93,dellavalle94}.

After more than twenty years in quiescence, \grs~was observed again in outburst by {\it MAXI} \citep{negoro16,masumitsu16}. \cite{bassi19} reported on the \grs~outburst monitoring, performed both in the radio band with the Australia Telescope Compact Array (ATCA), the Karl G.\ Jansky Very Large Array (VLA) and the Australian Long Baseline Array (LBA), and in X-rays with the XRT and BAT telescopes on board the {\it Neil Gehrels Swift Observatory} (hereafter {\it Swift}).
During this outburst (lasting about one year), \grs~showed only the spectral characteristics of the hard and hard-intermediate states.
Recently, \cite{tao19} inferred a  disc inclination angle in the range $40^{\circ}$-$50^{\circ}$.
The radio observations indicated that the emission could originate in a compact jet, despite the accretion disc being possibly at the innermost stable circular orbit (ISCO), or with the hot accretion flow having re-condensed in an inner mini-disc \citep{bassi19}.  Moreover, it was observed that \grs~was located on the radio-quiet branch \citep[L$_{\rm R}\propto$L$_{X}^{1.4}$,][]{coriat11} on the radio/X-ray luminosity plane during the whole outburst. 

In this paper, we present results of our multi-wavelength campaign (from radio to $\gamma$-rays),  performed in 2017 February when the source was in a bright hard spectral state.
Our aim was understanding the nature of the soft $\gamma$-ray emission ($\geq$200\,keV), also observed in other sources. We detected this emission with the SPI telescope and applied different models to investigate its origin.

\section{Observations and data reduction}\label{obs}

On 2017 February 9, when \grs~was in a bright HS, a multi-wavelength campaign was performed offering a sampling of observations from radio to $\gamma$-rays. 
The source was simultaneously observed with \integral, \swift, the Rapid Eye Mount Telescope (REM) and the ATCA.  
The ATCA data reduction has already been reported in \cite{bassi19}, where we measured flux densities of $3.28\pm0.05$\,mJy at 5.5\,GHz and of $3.04\pm0.03$\,mJy at 9\,GHz. The radio spectral index $\alpha=$ -0.15$\pm$0.08 was consistent with a flat-spectrum compact jet \citep{bassi19}. 

\subsection{REM near-IR Observations}
Observations in the near-IR filters $\mathrm{J, H, K}$ were obtained with the 60-cm robotic telescope REM located at the ESO-La Silla Observatory, and equipped with the IR camera REMIR \citep{vitali03}. The instrument has a field-of-view of $\sim 10^{\prime}\times 10^{\prime}$ and a 512-pixel camera with a pixel scale of $\mathrm{1.1^{\prime\prime}/pix}$.  
Observations and preliminary reductions and calibrations are done in a completely automated way by the robotic system using the pipeline AQuA \citep[Automatic QUick Analysis,][]{testa04}, and pre-processed images and initial catalogues are archived and then distributed to the program PIs. 
Observations of \grs~were performed during the night of 2017 February 9, by acquiring one single observation in the $\mathrm{J}$ and $\mathrm{H}$ filters, and two points in the $\mathrm{K}$ filter. The acquisition strategy is based on a series of frames acquired by rotating a filter wedge along the optical path in order to obtain five displaced images that are then combined together to obtain an ``empty sky" image of the field by median filtering the single images, that is then subtracted from the original frames. The sky-subtracted, flat-fielded frames are then registered together and summed to obtain the final science image. By using this pipeline, the final images have exposure times of 300\,s, 150\,s, and 75\,s for the $\mathrm{J, H}$ and $\mathrm{K}$ filters, respectively. 
The final science frames were reduced and analysed using PSF--fitting photometry package DAOPHOT \citep{stetson89,stetson94} and calibrated with the 2MASS survey by matching the 2MASS catalogue with the output DAOPHOT list after transforming the image coordinates into RA, Dec couples by using the astrometric information available in the header and obtained during the observation from the automatic REM pipeline. The single measurements for each night and for each filter were then matched together and the final results are shown in Table \ref{tab:uvot_rem}.
\begin{table*}
	\centering
	\caption{REM  and \swift/UVOT start time of the observations analysed in this work, in Terrestrial Time (TT) and MJD. In the columns are reported the exposure time and the filters used. In the last column are reported the REM magnitudes and the UVOT flux density in units of $10^{-16}$\,erg\,s$^{-1}$\,cm$^{-2}$\,\AA$^{-1}$ for each filter.}
	\label{tab:uvot_rem}
	\begin{tabular}{ccccc} 
		\hline
		Date & MJD & Exposure & Filter & Magnitude/Flux\\
        (TT)& & (s) &  &   \\
        \hline
        2017-02-09T07:47:00.384 &57793.32466 &300.0  &$\mathrm{J}$& 14.18$\pm$0.22 \\     	
        2017-02-09T07:52:33.542 &57793.32834 &150.0  &$\mathrm{H}$& 13.81$\pm$0.14  \\    	
        2017-02-09T07:56:11.098 &57793.33077 &75.0  &$\mathrm{K}$& 13.84$\pm$0.29   \\   	
        2017-02-09T07:58:12.403 &57793.33218 &75.0  &$\mathrm{K}$& 13.59$\pm$0.16 \\
        \hline
        2017-02-09T18:11:07     &57793.76&109&$\mathrm{U}$ & 4.8$\pm$ 0.5\\
                                &        &109&$\mathrm{B}$ & 6.5$\pm$ 0.6\\
                                &        &109&$\mathrm{V}$ & 9.3$\pm$ 1.0\\
                                &        &217&$\mathrm{UW1}$ & 1.68$\pm$0.34\\
                                &        &436&$\mathrm{UW2}$ & < 1.19	\\
                                &        &339&$\mathrm{UV2}$ & < 1.46\\
        \hline
	\end{tabular}
\end{table*}

\subsection{Swift Observations}\label{XRT}
On 9 February 2017, six observations were performed with the X-Ray Telescope \citep[XRT,][]{burrows05} and the Ultraviolet/Optical Telescope \citep[UVOT,][]{roming05} on board \swift~\citep{gehrels04}. 
The data were processed using the {\sc ftools} software package in HEASoft v.6.26 and the \swift\ relative Calibration Database (CALDB).
The criteria adopted for the pile-up correction and the source spectrum extraction were performed as described in \cite{bassi19}.
To apply the $\chi^2$ statistics, the energy channels were grouped to have at least 50 counts per energy bin. 
The average XRT spectrum obtained from the six observations performed was strongly affected by the known strong instrumental silicon (1.84\,keV) and gold (2.2\,keV) edges\footnote{\url{https://heasarc.gsfc.nasa.gov/docs/heasarc/caldb/swift/docs/xrt/SWIFT-XRT-CALDB-09\_v19.pdf}}. 
Therefore, we decided to use only one of the six spectra of \grs~in the spectral analysis described in the next sections. 
We have made sure that the spectral parameters and the fluxes of the 6 observations were consistent with each other, and we selected the pointing 00034924012.

By combining the 6 XRT observations taken on February 9 we extracted and averaged the power density spectrum (PDS), using custom software written in IDL\footnote{GHATS,\url{http://www.brera.inaf.it/utenti/belloni/GHATS\_Package/Home.html}}.
We used $\approx$29-s long intervals and a Nyquist frequency of $\approx$64\,Hz, and from each interval we computed a PDS in the energy band 0.4--10.0\,keV. We then averaged the PDSs and we normalised the result according to rms$^2$/Hz normalisation.

Furthermore, we determined magnitudes and fluxes for UVOT images using the task {\it uvotsource}, selecting a circular region of  5\arcsec~at the best source coordinates and a larger region with no other source as background. The flux densities calculated in the different bands for each observation are reported in Table \ref{tab:uvot_rem}. 

From 2016 December 1 (MJD 57723), \grs~was observed almost daily in survey mode with the Burst Alert Telescope \citep[BAT,][]{barthelmy05}. 
To increase the statistics of the BAT spectrum, we selected the \grs~observations (available from the HEASARC public archive) performed   from 2017 February 2 (MJD 57786) to March 15 (MJD 57827), when the source was constant in flux and  spectral shape in hard X-rays \citep{bassi19}.
The data were processed using the \texttt{BAT-IMAGER} software \citep{segreto10}. 
We extracted the spectrum in 29 channels with logarithmic binning in the energy range 15--185\,keV. The official BAT spectral redistribution matrix was used. \\

\subsection{INTEGRAL Observations}
We collected and analysed the \integral~\citep{winkler03} ToO campaign data performed on February 9 and all public observations of \grs~from 2017 February 2 to March 15 (from revolution 1780 up to 1793). \\
The SPI \citep{vedrenne03} analysis was performed using the SPI Data Analysis Interface (SPIDAI) tools\footnote{\url{http://sigma-2.cesr.fr/integral/spidai}}. The SPI camera records the signal contribution from the sources in the field of view (FOV) plus the background. To correctly describe the data, we introduced some information on these components.
We assumed a constant background during 12 Science Windows (SCW).
The sky models were determined based on the IBIS/ISGRI map of each SCW within the revolution and of the full revolution. We selected the most significant sources ($\sigma\geq$10) present in the SPI FOV during each revolution. The sources' variability was defined by their IBIS/ISGRI light curves within the SCW binning time (1 SCW$\sim$1\,hr). 

The \grs~SPI spectrum for each revolution was extracted in 39 channels in the energy range 25--1000\,keV, requiring at least 2$\sigma$ significance in the higher energy bin (i.e. $\geq$300\,keV). 

\section{Broad band X/gamma-ray spectral modelling}
The spectral variability study based on XRT and BAT data has been reported by \cite{bassi19}. The authors did not observe any significant variability in the day-averaged emission of the source up to the 150\,keV energy band from MJD 57786 to MJD 57827.\\
To investigate the variability above 150\,keV, we fitted the SPI spectrum of each revolution  with a {\it cutoffpl} model. We observed that the photon index and high energy cut-off values were consistent within the errors and the flux variation in the 25--300\,keV energy band was lower than about 20 per cent.
Moreover, we did not observe any significant spectral variability in the energy range 200--600\,keV: i.e.\ the hardness ratio [300--600]\,keV/[200--300]\,keV (with a revolution time bin) is constant.\\ 
Therefore, it was possible to use the averaged BAT and SPI spectra from MJD 57786 to MJD 57827 (February 2 to March 15, 2017) with the XRT pointing (MJD 57793.76) to perform a broad band spectral analysis in the largest energy range possible. 

The spectra were fitted with {\sc xspec} v. 12.9.1p. All the errors reported are at the 90 per cent confidence level.
A systematic error of 2 per cent was introduced in all the broad band spectra.
In the fits we adopted the cosmic abundances of \cite{wilms00} and the cross-sections of \cite{verner96}.\\
Even though we obtained a good $\chi^{2}$ ($\chi^{2}_{\nu}$(dof)=1.05(336)) applying an absorbed thermal Comptonisation model ({\it tbabs} plus {\it nthcomp} in {\sc xspec}), we observed significant residuals above 200\,keV. The addition of a power-law component at high energy allows us to eliminate these residuals and improve the fit ($\chi^{2}_{\nu}$(dof)=0.97(334)), with a F-test probability 2.32$\times$10$^{-6}$.\\
We found a hydrogen column density N$_{\rm H}$=0.70$\pm0.02\times$10$^{22}$\,\cm, a photon index $\Gamma$=1.68$\pm0.01$ and an electron temperature \kte=50$^{+4}_{-3}$\,keV, consistent with the values reported in \cite{bassi19}.
Then, we observed that a power-law with photon index $\Gamma$=1.12$^{+0.21}_{-0.63}$ reproduces the high energy excess above 200\,keV and a flux of F$_{[200-600]\,{\rm keV}}$= 2.5$\times10^{-9}\,\rm erg\,cm^{-2}\,s^{-1}$, was obtained.

\subsection{Hybrid Comptonisation models} \label{hybrid}

In the following we present the results of the broad band spectral analysis using the physical hybrid thermal/non-thermal Comptonisation models: i.e.  {\sc eqpair} \citep{coppi99} and {\sc belm} \citep{belmont08}.

\subsubsection{Unmagnetised model}
\begin{figure}
	\includegraphics[width=9.5cm]{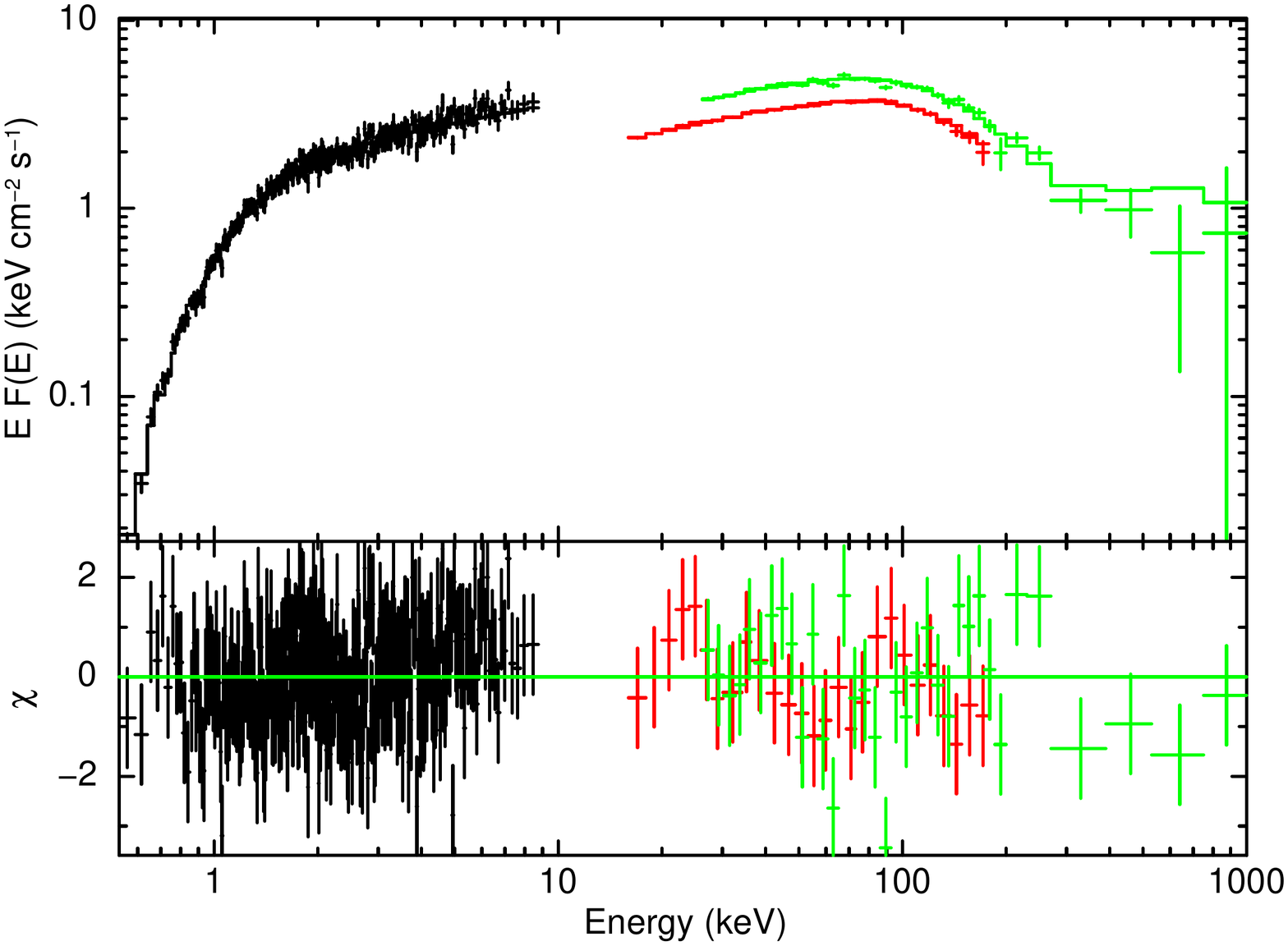}
    \caption{Broadband energy spectra of XRT pointing $\#12$ (black points) with average BAT (red points) and SPI (green points) spectra in the simultaneous time interval (57792.968--57793.975 MJD) fitted with the {\sc eqpair} model.}
    \label{fig:eqpair}
\end{figure}
\begin{table*}
\centering
    \caption{The best-fit parameters of the broadband XRT, BAT and SPI spectrum performed with an absorbed thermal/non-thermal Comptonisation ({\sc eqpair}) and {\sc belm} models. (1) the ratio of the total power provided to the electrons of the plasma to the soft radiation coming from the accretion disc and entering in the corona, (2) the ratio of the fraction of power used to accelerate non-thermal electrons to the total power supplied to the plasma (frozen to unity in the case of the {\sc belm} model); (3) optical depth of ionized electrons and  (4) total Thomson optical depth, (5) hydrogen column density in units of $10^{22}$\,cm$^{-2}$, (6) Comptonising electron temperature in keV, (7) slope of the power law distribution of the injected non-thermal electrons, (8) the magnetic compactness over the fraction of the power used to accelerate non-thermal electrons, (9) unabsorbed bolometric flux (0.1-1000\,keV) and (10) the reduced $\chi^{2}$.}
    \label{tab:eqpair}
    \begin{tabular}{lcccccccccccc}
    \hline
    &\lh/\ls & \lnth/\lh & $\tau_{\rm p}$ &  $\tau_{\rm T}$ &  N$_{\rm H}$  & \kte & $\Gamma_{inj}$ &\lb/\lnth & Flux   & $\chi^{2}_{r}$(dof) \\
    &        &           &                &                 &  (10$^{22}$\,\cm) & (keV) &   &&(\ergcm)&\\
    &(1)&  (2) & (3) & (4) & (5) & (6)&(7) &(8)&(9)& (10)\tabularnewline
    \hline
        \tabularnewline
     {\sc eqpair} &9.69$^{+0.31}_{-0.21}$ & 0.57$^{+0.04}_{-0.01}$ & 1.79$^{+0.08}_{-0.03}$ &  $\sim$2.6 &0.71$\pm0.01$ & $\sim$30& (2.5)&-&4.5$\times10^{-8}$& 1.03(334)\\
     {\sc belm}   &- & 1  & - &  3.3$\pm0.1$ & 0.70$\pm0.02$& $\sim$34 &2.64$^{+0.20}_{-0.14}$&0.017$^{+0.016}_{-0.006}$&4.6$\times10^{-8}$& 0.94(341)\\
    \tabularnewline
    \hline
    \end{tabular}
\end{table*}

First, we fit the broad band spectrum with the hybrid thermal/non-thermal model {\sc eqpair} \citep{coppi99}. 
In this model the emission of the disc/corona system is assumed to arise from a spherical, homogeneous, isotropic, hot ionised plasma cloud with continuous injection of relativistic electrons illuminated by soft photons emitted by the accretion disc. The model takes into consideration Compton scattering, e$^{\pm}$ pair production and annihilation, Coulomb interactions and bremsstrahlung processes.
The electron distribution at low energies is Maxwellian, with an electron temperature \kte, whereas at high energies the electrons are characterised by a non-thermal distribution.
The properties of the plasma are defined by the non-dimensional compactness parameter
\begin{equation*}
    l=\frac{\sigma_{\rm T}}{m_{e}c^{3}}\frac{L}{R} 
\end{equation*}
where $\sigma_{\rm T}$ is the Thomson cross-section, m$_{e}$ the electron mass, $c$ the speed of light,  L is the total power of the source supplied to soft seed photons and electrons, and $R$ is the radius of the emitting region. For a detailed description of the model parameters we refer to \cite{coppi99}.
Since the source was in a bright HS \citep[see Figure 2 in][]{bassi19}, most of the luminosity comes from the corona and the contribution of the accretion disc would be small, or negligible.
Following the assumption made in previous  spectral analyses of \Cyg~and \gx~ \citep[][respectively]{gierlinski99,delsanto08} we fixed the soft photon compactness \ls, which is proportional to the luminosity of the thermal disc radiation  entering the corona, to \ls=10. 
It is worth noticing that the predicted spectral shape is not sensitive to \ls~but depends mostly on the compactness ratios,
\begin{itemize}
    \item[-] \lh/\ls, where \lh~is proportional to the total power provided to the electrons of the plasma; 
    \item[-] \lnth/\lh, the fraction of the total power used to accelerate non-thermal electrons over the total power supplied to the plasma, (i.e. including also the thermal heating of the  electrons \lth, \lh=\lnth+\lth).
\end{itemize}
Since the soft component is weak, constraining the seed photon temperature (kT$_{\rm max}$) makes the fit unstable, so we kept this parameter frozen at a typical value observed in this state, i.e. 0.3\,keV.
The non-thermal electrons are injected with a power law distribution $\gamma^{-\Gamma_{inj}}$, with a Lorentz factor  $\gamma$ in the range 1.3-1000. Because of the degeneracy of the parameters, following \cite{delsanto16}, we fixed the slope of the electron distribution at the value $\Gamma_{inj}$=2.5 expected in shock acceleration models.
Introducing a reflection component in the spectral model did not improve the fit significantly, therefore we fixed the reflection amplitude ($\Omega/2\pi$) at zero.

In Figure \ref{fig:eqpair}, we show the best-fit model of the broad band spectrum. The best-fit parameters are reported in Table \ref{tab:eqpair}.

The best-fit hydrogen column density is compatible with the value obtained by \cite{bassi19}. 
The \lh/\ls~values are close to the values measured for other 
BHBs in bright hard and intermediate states: i.e. \gx~\citep{delsanto08}, \Cyg~\citep{delsanto13} and Swift J174510.8--262411 \citep{delsanto16}. 
The best-fit parameters suggest that about half of the total power is supplied to the plasma in the form of non-thermal electrons (see Tab. \ref{tab:eqpair}). 

\subsubsection{Magnetised model}
\begin{figure}
    \includegraphics[width=\columnwidth]{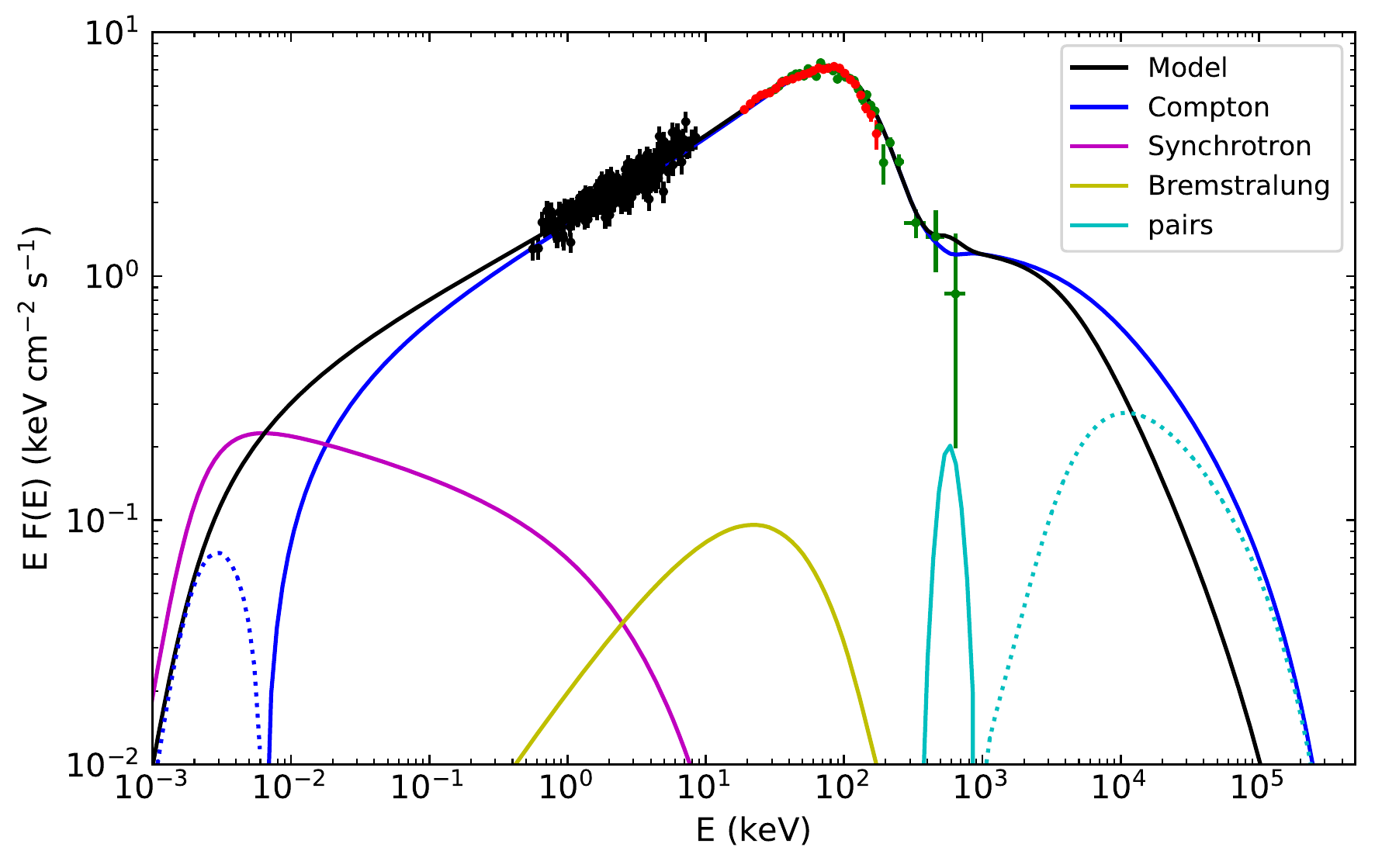}
    \\
    \includegraphics[width=\columnwidth]{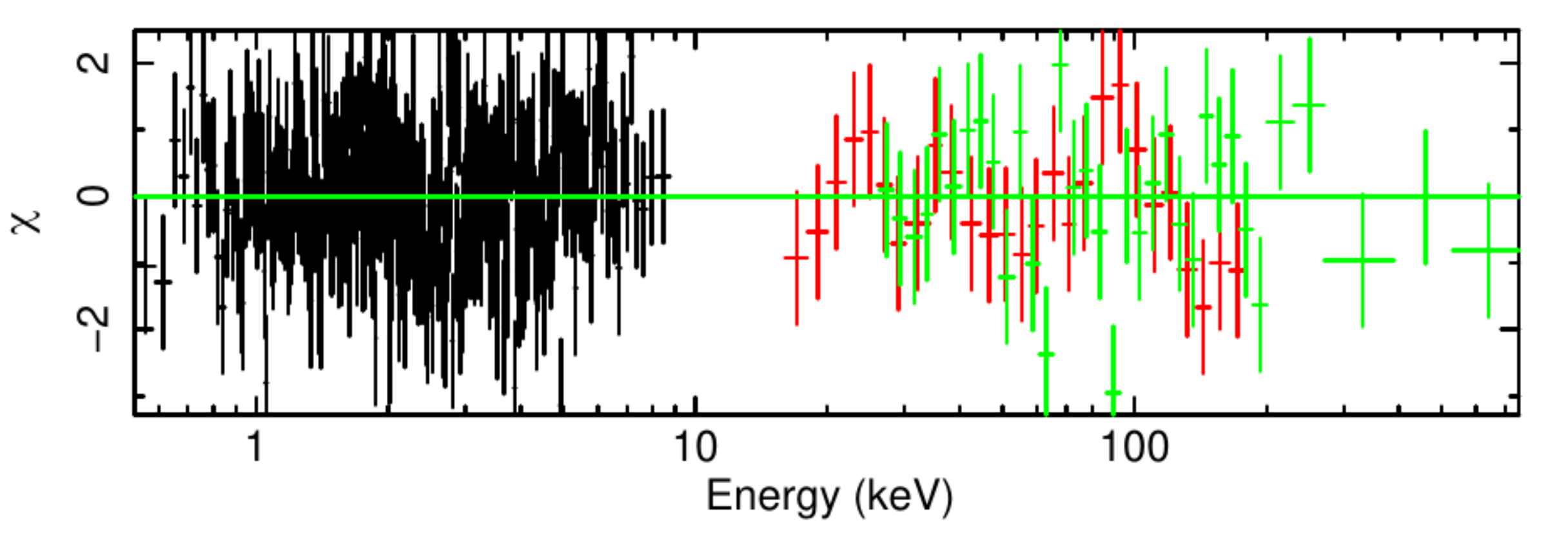}
    \caption{{\it Top:} Broadband energy spectra of XRT pointing $\#12$ (black points) with average BAT (red points) and SPI (green points) spectra in the simultaneous time interval (57792.968--57793.975 MJD) fitted with the {\sc belm} model. The total model (black line) corresponds to the fit of the broadband spectrum. The positive (solid lines) and negative (dashed lines) components' contributions to the spectrum: Compton (blue), Synchrotron (magenta), Bremsstrahlung (yellow) and pair annihilation/production (cyan). {\it Bottom:} Residuals obtained with the magnetised model {\sc belm} by fitting XRT, BAT and SPI spectra.}
    \label{fig:belm}
\end{figure}
The second hybrid thermal/non-thermal Comptonisation model adopted to fit the data of \grs~is {\sc belm} \citep{belmont08}. 
This model, in addition to the processes considered in {\sc eqpair}, takes also into account  the magnetic field in the corona and the self-absorbed synchrotron emission from the energetic leptons of the Comptonising plasma. The effects of magnetic field are quantified through the magnetic compactness parameter: 
\begin{equation}\label{eq:compactness_B}
    l_{B}=\frac{\sigma_{\rm T}}{m_{e}c^{3}}R\frac{B^2}{8\pi} 
\end{equation}
Power is provided to the system in the form of  a continuous  injection of electrons in the corona with a  power-law energy distribution $\gamma^{-\Gamma_{inj}}$ ($1<\gamma<1000$). At the same time electrons are removed form the overall distribution to ensure that the number of particles is conserved and mimic a non-thermal acceleration process.
\cite{malzac09} showed that particles accelerated through non-thermal mechanisms can be thermalised efficiently on times scales shorter than the light crossing time of the corona under the effects of the synchrotron boiler \citep{haardt94,ghisellini98} as well as Coulomb collisions. So that, in the end the equilibrium distribution is an hybrid distribution similar to that assumed in {\sc eqpair}.
However, unlike {\sc eqpair} where the electron energy distribution at low energies is assumed to be Maxwellian, in {\sc belm} the thermalisation process is treated self-consistently. 

We computed a table for a pure Synchrotron Self Compton (SSC) model, where the protons are cold (\lc=0), all the power is injected in the form of non-thermal particles, and the external photons (i.e. from the disc) are negligible (\ls=0). 
Also in this model the spectral shape depends mostly on the ratio \lb/\lnth~and is relatively insensitive to the value of the individual parameters. 
Under these conditions, the soft, self-absorbed synchrotron emission produced by the interaction between the non-thermal electrons and the magnetic field ($B$) peaks around a few eV. This component is then Compton up-scattered by the hybrid thermal/non-thermal electron distribution, extending the spectrum up to the X/$\gamma$-ray energies.\\
The broad band spectrum and the different components' contributions to the model are shown in Figure \ref{fig:belm} and the best-fit parameters are reported in Table \ref{tab:eqpair}.
The hydrogen column density and the electron temperature are compatible with the values obtained with {\sc eqpair} (N$_{\rm H}$=0.70$\times$10$^{22}$\,\cm~and  \kte$\sim$34\,keV, respectively). 
In the {\sc belm} model, the electron acceleration index is better constrained, so we leave it free to vary, obtaining the best-fit value $\Gamma_{inj}$=2.64. Then, we found a total Thompson optical depth of 3.3. 

Since the spectral shape is determined by the ratio \lb/\lnth, the best-fit value of \lb~is highly dependent on our choice of \lnth~and \lb~does not represent the true magnetic compactness of the source. In order to estimate the observed magnetic field we need to rescale \lb~to the real source compactness \los:
\begin{equation}
    l_{\rm obs}=\frac{4\pi D^2 F \sigma_{\rm T}}{R m_{e} c^3}
\end{equation}
estimated from the source luminosity. The observed magnetic field compactness can be expressed as \citep[see][]{delsanto13}: 
\begin{equation}\label{eq:lb_oss}
    l_{B,\rm obs}=\frac{l_{B}}{l_{\rm nth}} \frac{l_{\rm nth}}{l} l_{\rm obs}
\end{equation}
It is worth noticing that $l$=\lnth~since we are considering a case of pure SSC and purely non-thermal heating of the leptons. 
Using the equations \ref{eq:compactness_B} and \ref{eq:lb_oss} we can estimate the magnetic field $B$ in the limit of pure SSC emission.
In the case where in addition to the synchrotron photons, soft photons from the disc represent a significant source of seed photons for the Comptonisation process, the efficiency of Compton cooling of the electrons in the corona is increased. Therefore, every other parameter being equal, the equilibrium temperature of the thermal electrons is lower and a steeper Comptonisation spectrum with a lower energy cut-off is expected. In order to keep the model in agreement with the observed spectral shape, the Compton cooling needs to be reduced by cutting the amount of synchrotron seed photons. This can be achieved only by reducing the magnetic field. Therefore any model combining both disc and synchrotron photons would require a magnetic field that is lower than the one given by the equations \ref{eq:compactness_B} and \ref{eq:lb_oss}.\\
Considering a pure SSC model, assuming a corona size of $R\sim$20\,R$_{\rm g}$ \citep{delsanto13} for a $M_{\rm BH}=4.9\,M_{\odot}$  and using the estimated bolometric flux F=$4.6\times10^{-8}\,\rm erg\,cm^{-2}\,s^{-1}$ and the best-fit value \lb/\lnth=0.017 (see Tab. \ref{tab:eqpair}), we calculated an upper limit on the magnetic field in the corona  $\rm B<1.5\times10^6\left(\frac{20\,\rm R_{\rm g}}{\rm R}\right)$\,G.\\

The magnetic field over the magnetic field in equipartition can be expressed in terms of compactness as:
\begin{equation*}
    \frac{l_{B}}{l_{B_{R}}} = \frac{l_{B}}{l_{nth}} \frac{l_{nth}}{l} \frac{\frac{4\pi}{3}}{1+\frac{\tau_{T}}{3}}
\end{equation*}
where $l_{B_{R}}$ expresses in terms of compactness the equipartition between the magnetic field and the radiative energy density in the limit of low energy photons \citep[h$\nu$<m$_{e}c^{2}$,][]{lightman87}.
Using this equation and the best-fit parameters we calculated \lb/$l_{B_{R}}$=0.03 for our source.
Therefore the magnetic energy density represents only a few percent at most of the radiation energy density in the corona. 

\section{Spectral energy distribution}

\subsection{The Internal Shock Emission Jet model} \label{jet}

We used the internal shock emission model \citep[{\sc ishem},][]{malzac13,malzac14} to investigate whether the soft $\gamma$-ray emission (above 200\,keV) detected in \grs~by SPI can be explained as synchrotron emission from the jets.

The {\sc ishem} model simulates the hierarchical merging of shells of plasma ejected at the base of the jet with variable velocities. A fraction of the kinetic energy of these ejecta is converted into internal energy and radiation by the shocks produced when shells of different velocities collide.
The fluctuation of the Lorentz factor $\Gamma_{j}$ of the shells in the jet is defined assuming that its power spectrum has the same shape as the PDS observed in X-ray. 
In presence of non-thermal acceleration processes in the jet, the synchrotron emission of the non-thermal particles generally dominates over that of the lower energy thermalised component \citep{wardzinski01}.
For this reason, the emission from the Maxwellian component of the electron energy distribution is neglected in {\sc ishem}. 
We note that some authors have considered jet emission models involving only thermal relativistic electron distributions \citep{peer09,tsouros17}.

The  model takes as input a number of parameters related to the properties of the source, the jet and the distribution of the radiating particles. In particular, for \grs, we used a distance of 2.4\,kpc \citep{dellaValle93} and a black hole mass of 4.9\,M$_{\odot}$ \citep{masetti96}.
Following \cite{malzac14}, we defined the ejecta assuming a relativistic flow with an adiabatic index of 4/3 and a volume filling factor of 0.7. Furthermore, the jet shells are ejected from a region at the base of the jet with typical radius of 10\,R$_{\rm g}$ \citep{malzac14} and with a time-averaged bulk Lorentz factor $\Gamma_{av}$ which is a free parameter of the model.
The jet half-opening angle $\phi$, the jet power \pj\ and the jet inclination angle $\vartheta$ are also free to vary.
The electrons responsible for the synchrotron emission follow a power law energy distribution extending in the typical XRB range from $\gamma_{min}$=1 to $\gamma_{max}$=10$^{6}$ \citep{malzac14}, with an index of the electrons' energy distribution $p$.\\
Given a set of parameters we can compute a synthetic Spectral Energy Distribution (SED) using the associated X-ray PSD. 
However the shape of the SED is determined mostly by the power spectrum of the Lorentz factor fluctuations. The only other parameter that has an effect on the SED shape is the slope $p$ of the power law energy distribution of the accelerated electrons. Indeed $p$ affects the slope of the optically thin synchrotron emission at frequencies above the jet spectral break. In particular the $p$ parameter will control the jet luminosity around 1\,MeV and determine whether the jet is responsible for the observed emission above 200\,keV. In the following we show the results of spectral fits with $p$ frozen to two different values: $p=2.5$ and $p=2.1$ which are both in the range of values expected in shock acceleration.\\
The effect of the other parameters is to change the normalisation or shift the SED along the frequency axis without altering the SED profile. If the synthetic SED shape does not fit the data there is no way to fine tune these parameters in order to match the data. On the other hand if the predicted shape of the SED is similar to the observed SED it is difficult to constrain the model parameters, as there is a strong degeneracy that makes it possible to reproduce the observed SED with many different parameter combinations. A detailed description on the behaviour of the model parameters is reported by \cite{peault19}.\\

\subsection{Fitting procedure}

    \begin{figure}
    \centering
	\includegraphics[width=9.5cm]{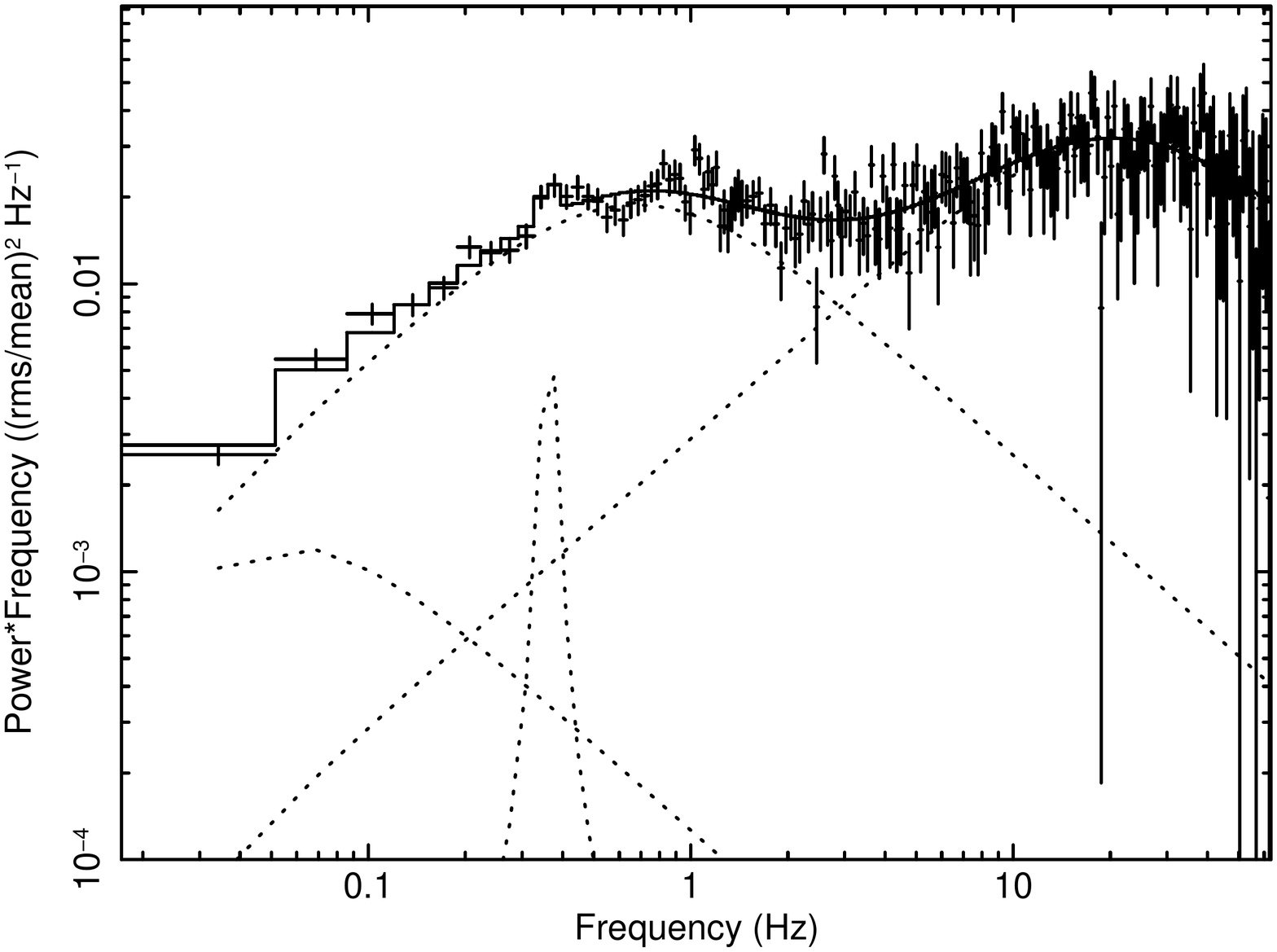}
    \caption[Average XRT PDS of the six pointings on 9 February 2018.]{Average X-ray PDS of the six XRT pointings on 9 February 2018. They were fitted with three Lorentzians, of which two are zero-centred, plus a QPO. }
    \label{fig:pds}
\end{figure}

\begin{table}
\centering
\caption[Results of the average XRT PDS for \grs, fitted with four  Lorentzians.]{Results of the \grs~average XRT PDS fitted with multiple Lorentzians, given individually by $P(\nu)=r^2/\pi [\Delta^2+(\nu -\nu_{0})^2]$ \citep{belloni02}. Values show the integrated \rms~over the full range of $+\infty$ to $-\infty$ ($r$), the centred frequency $\nu_{0}$ and its Half Width at Half Maximum $\Delta$. }
    \begin{tabular}{ccccc}
    \hline
    Lorentzian &$r$ & $\nu_{0}$ & $\Delta$\\
    Component&  & &\\
    && (Hz) & (Hz)\\
    \hline
        \tabularnewline
     1& 0.1$^{+0.01}_{-0.01}$ & (0)                      & (0.05)                    \\
     2& 12.7$^{+0.9}_{-0.9}$ & $<2.9$                  & 20.9$^{+2.7}_{-3.1}$ \\
     3& 1.80$^{+0.09}_{-0.10}$ & (0)                       & 0.68$^{+0.07}_{-0.06}$  \\
     4& 0.03$^{+0.03}_{-0.06}$ & 0.36$^{+0.01}_{-0.01}$  & 0.01$^{+0.05}_{-0.02}$  \\
     \tabularnewline
    \hline
    \end{tabular}
    \label{tab:qpo}
\end{table}
The fit of the observed SED of \grs~was performed by following the steps below:
\begin{itemize}
    \item {\it Compute the synthetic SED.}  In order to generate the synthetic jet SED we used the best-fit parameters obtained for the average X-ray PDS in {\sc ishem} and an initial set of parameters related to the source, jet and radiating particles' properties (see Section \ref{jet}).\\
    In Figure \ref{fig:pds} we show the total PDS fitted with three Lorentzians, two of which are zero-centred. A fourth Lorentzian was used to fit a QPO at 0.36\,Hz (2.9$\sigma$ significance).
    The best-fit ($\chi^{2}_{r}$(dof)=1.01(178)) parameters are reported in Table \ref{tab:qpo}.
    It is worth noting that if the X-ray QPO found in the PDS of \grs~(see Fig. \ref{fig:pds}) is a geometrical effect due to the accretion disc precession \citep{motta15}, it is not related to the intrinsic aperiodic variability of the accretion flow. 
    Therefore, it should not be introduced in the model to simulate the shells' propagation. However the QPO is weak  and its effect on the synthetic SED is negligible. Indeed, performing {\sc ishem} simulations with and without the QPO shows that the resulting SEDs differ by no more than $4\%$. Such small differences do not affect the analysis. Therefore, in the next Section (\ref{results}) we will show the results obtained including all the four PDS components in {\sc ishem}.\\
    \item {\it Fit with the synthetic SED.} Once we obtained the synthetic SED, we fitted the multi-wavelength data of \grs~using the {\sc xspec} local jet model named {\sc ish} with the break frequency and the break flux density as free parameters. This allowed us to compare the synthetic SED with the observed data and to quantify the shift in terms of frequency ($\Delta \nu$) and flux density between the synthetic and the observed SED.\\
    \item {\it Define the reasonable parameter combinations.} 
    For a BHB the flux normalisation and the frequency break scale as  \citep{malzac13} :
        \begin{equation}\label{eq:flux_ishem}
    F_{\nu}\propto\frac{\delta^{2}i_{\gamma}^{\frac{5}{p+4}}}{D^{2}\tan\phi}\left(\frac{P_{jet}}{(\Gamma_{av}+1)\Gamma_{av}\beta}\right)^{\frac{2p+13}{2p+8}};
    \end{equation}
        \begin{equation}\label{eq:breac_frequency}
    \nu_{b}\propto \frac{\delta i_{\gamma}^{\frac{2}{p+4}}}{\tan\phi}\left(\frac{P_{jet}^{\frac{p+6}{2p+8}}}{(\Gamma_{av}+1)\Gamma_{av}\beta}\right)^{\frac{3p+14}{2p+8}},
    \end{equation}
    where $i_{\gamma}=(2-p)(\gamma_{max}^{2-p}-\gamma_{min}^{2-p})^{-1}$, $\beta=\sqrt{1-\Gamma_{av}^{-2}}$, \newline $\delta=[\Gamma_{av}(1-\beta\cos{\vartheta})]^{-1}$ and $D$ is the source distance.\\
    Based on the best-fit normalisation and frequency shift parameters that we found with the {\sc ish} model, we defined a new combination of physical parameters that produced the required shift in frequency and normalisation with respect to the initial model  \citep{malzac13, malzac14, peault19}.

\end{itemize}

It is worth noting that in all the models, we assume that the optically thin synchrotron power law from the jet  extends up to at least 1\,MeV without any cooling break or high energy cut-off within the energy range covered by our observations.

\subsection{Accretion and ejection emission fitting results}\label{results}

\begin{table}
	\centering
	\caption{Spectral Energy Distribution of \grs. The fit was performed with an irradiated disc ({\it diskir}) plus jet internal shock emission model ({\sc ish}). In the table we report the best-fit values and the parameters of the simulation were obtained for a black hole of 4.9\,M$_{\odot}$ at 2.4\,kpc and assuming a jet with an electron distribution p=2.5 and p=2.1.}
	\label{tab:sed}
	\begin{tabular}{lll}
		\hline
		$p$                            & 2.1                      & 2.5 \\
		$\Delta\nu$                 & <$10^{-1.15}$                     & ($10^{-1.00}$)\\
		\hline
		N$_{\rm H}$ ($10^{22}$ \cm) & 0.65$^{+0.02}_{-0.02}$    & 0.67$^{+0.02}_{-0.02}$ \\
		E(B-V)                      & 0.95$^{+0.25}_{-0.02}$     & 1.05$^{+0.18}_{-0.20}$\\
		kT\_disk ($10^{-2}$\,keV)   & 4.11$^{+0.01}_{-0.02}$     & 5.29$^{+0.73}_{-0.77}$\\
		$\Gamma$                    & 1.67$^{+0.01}_{-0.01}$     & 1.68$^{+0.01}_{-0.01}$ \\
		\kte (keV)                  & 50$^{+4}_{-3}$             & 58$^{+3}_{-3}$\\
		L$_{c}$/L$_{d}$             & >2.02                      & 1.72$^{+3.03}_{-0.70}$\\
		fout ($\times 10^{-3}$)     & 1.66$^{+13.2}_{-1.5}$      & >0.1 \\
		logrout                     & 2.54$^{+0.24}_{-0.29}$     & 2.56$^{+0.31}_{-0.36}$\\
		$\chi$(dof)                 & 1.01 (425)                 & 1.06 (426)\\
				 & &\\
		$\vartheta$                 & 15                    &  15            \\
		$\Gamma_{av}$               & 4.5                    &  4.5           \\
		$\phi$                      & 10.6               &  6.0          \\
		\pj~ (L$_{\rm E}$)          & 0.04             &  0.05        \\
		\hline
	\end{tabular}
\end{table}

\begin{figure*}
	\includegraphics[width=17cm]{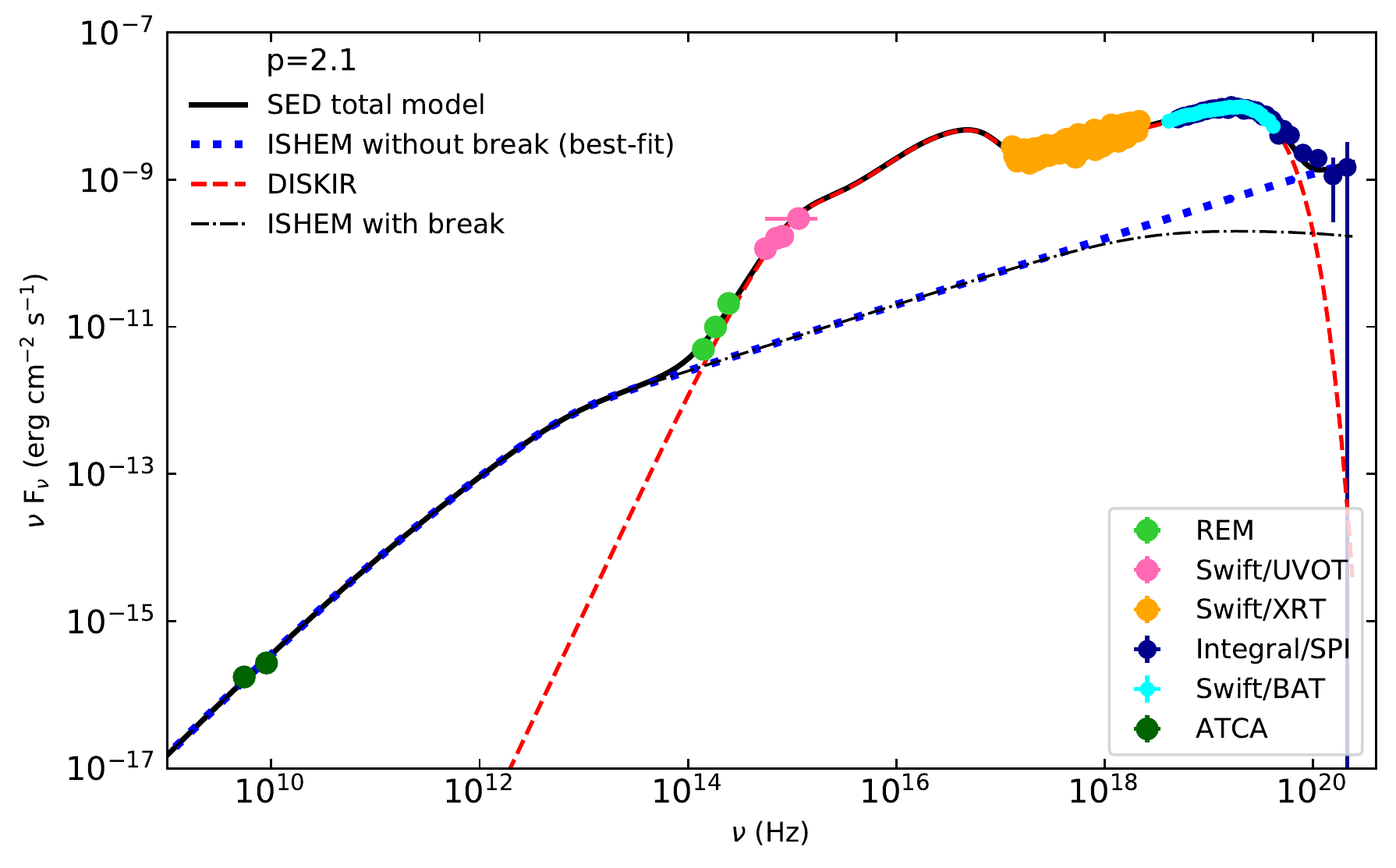}
	\\
    \includegraphics[width=17cm]{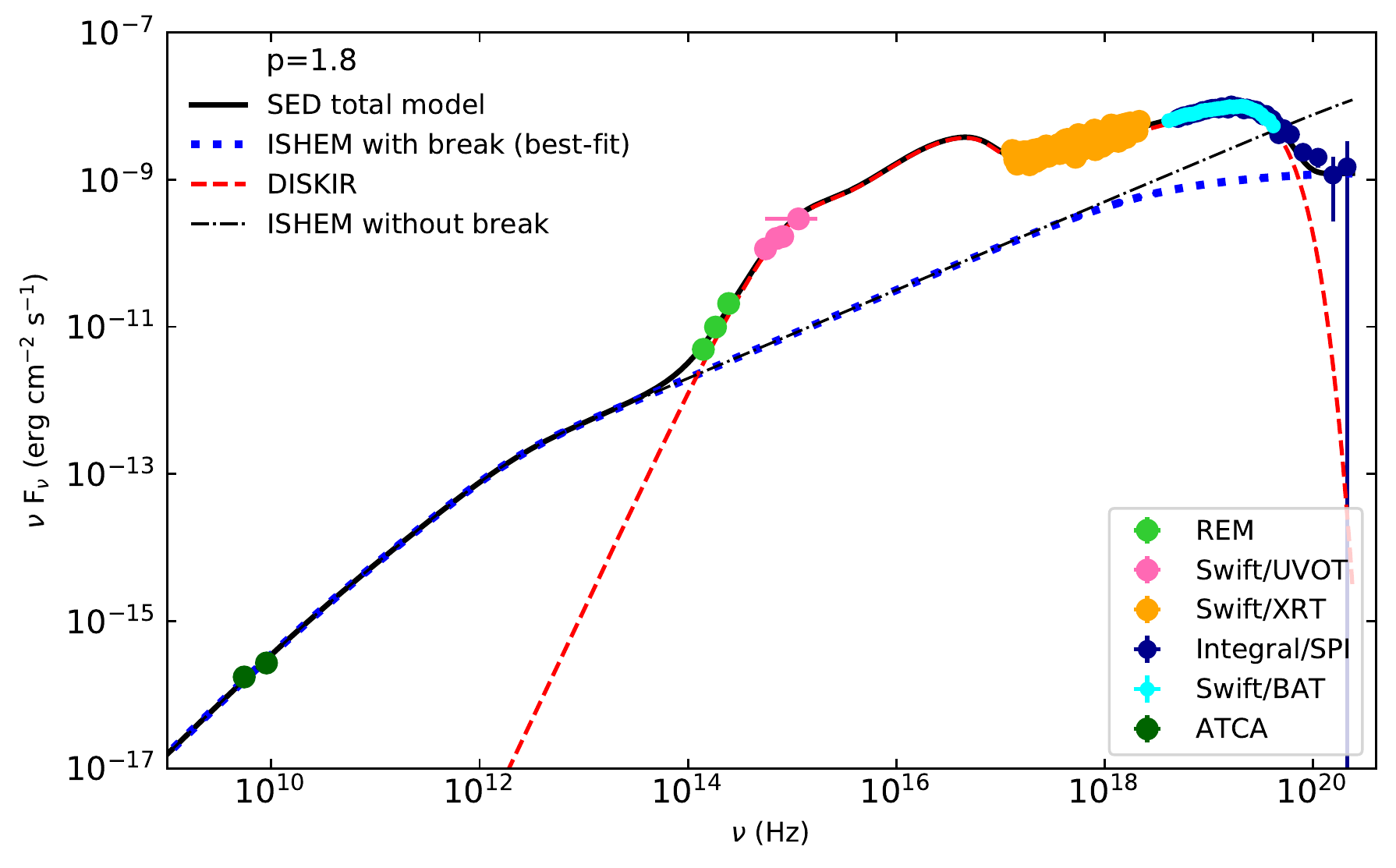}
    \caption{Spectral energy distribution of \grs~built with the data collected during the multi-wavelength campaign performed in February--March 2017. To reproduce the soft $\gamma$-ray emission observed (above $\sim$5 $\times$ $10^{19}$ Hz) we used the {\sc ishem} model assuming the electron distribution with p=2.1 (upper panel) and  p=1.8 (bottom panel). The accretion disc/corona contribution is modelled with the irradiated disc model {\sc diskir} (dashed red line). The blue dotted line represents, in both plots, the best-fit with the jet emission model {\sc ishem}. The black dash-dotted line shows the {\sc ishem} model with a break at 10\,keV when p=2.1 (upper panel) and the {\sc ishem} model without break when p=1.8 (bottom panel). The total best-fit model is shown as the solid black line.}
    \label{fig:sed}
\end{figure*}

We assembled the observed SED of \grs, from radio to $\gamma$-ray, from the data collected during the multi-wavelength campaign.\\
To fit the whole SED with {\sc xspec}, we created the spectra and the related response matrix from the ATCA, REM, and UVOT flux densities and magnitudes (see Section \ref{obs} and Table \ref{tab:uvot_rem}), by running the task {\it ftflx2xsp}.\\
To fit the SED,  the {\it ish} model was used together with the irradiated disc model {\it diskir} \citep{gierlinski08} to describe the contribution of the accretion flow emission and with the models {\it redden} \citep{cardelli89} and {\it tbabs} to take into account the IR/optical and the X-ray interstellar absorption, respectively. The {\it diskir} model accounts for the irradiation of the outer disc and the reprocessing of the X-ray photons in the optical/UV band, in addition to the thermal Comptonisation in a hot corona \citep{gierlinski08,gierlinski09}.\\
The best-fit parameters are shown in Table \ref{tab:sed}. Note that the goodness of our fits is mainly driven by the X-ray data.\\
The best-fit hydrogen column densities N$_{\rm H}$ are in agreement with the values obtained with our previous broad band spectral fits (see Section \ref{hybrid}). Moreover, the E(B-V) values obtained are in agreement with the value reported by \cite{dellavalle94} (<E(B-V)>=0.9$\pm$0.2).
The Comptonisation parameters $\Gamma$ and \kte~are similar to those typically found in hard states \citep[see e.g. ][ and references therein]{zdz04,done07}.
The ratio of luminosity in the Compton tail to that of the unilluminated disc (L$_{c}$/L$_{d}$) is higher than 1 as is expected in hard state spectra. The disc emission peaks in the UV band in agreement with what is expected for a disc truncated at large radii.
The outer disc radius in terms of the inner disc radius, is R$_{\rm out}\sim$10$^{2.5-2.6}$\,R$_{\rm in}$.\\

\subsubsection*{Jet parameters}\label{jet_parameter}

\begin{figure*}
	\includegraphics[width=17cm]{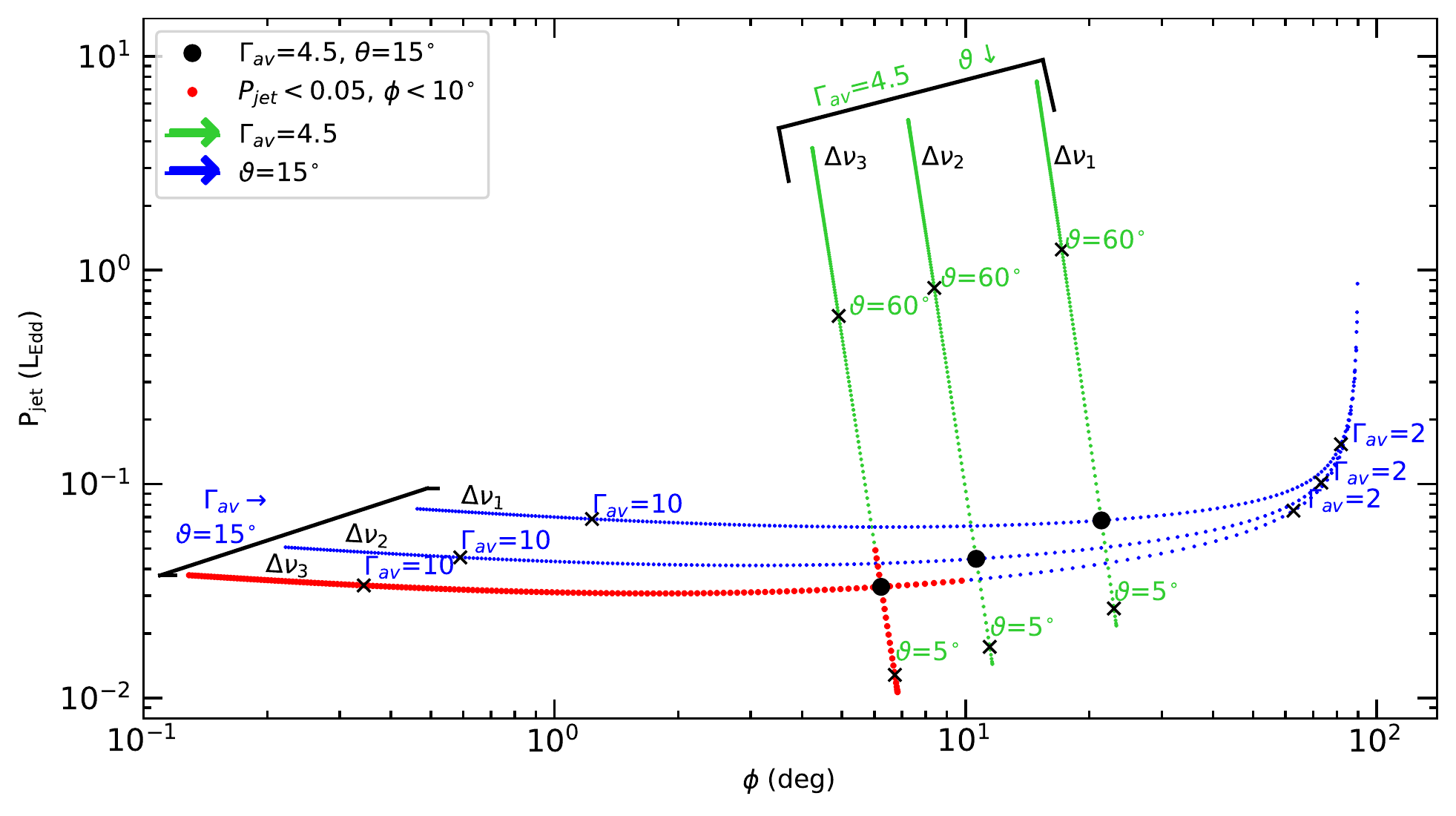}
    \caption{Evolution of the jet power (P$_{\rm jet}$) versus the jet opening angle ($\phi$) as a function of the mean Lorentz factor $\Gamma_{\rm av}$ and the jet inclination angle $\theta$ assuming $p=2.1$. We investigated the possible parameter combinations assuming three SED frequency shifts between the synthetic and observed SED: $\Delta \nu_{1}$=10$^{-1.5}$, $\Delta \nu_{2}$=10$^{-1.3}$ and $\Delta \nu_{3}$=10$^{-1.15}$. For each $\Delta \nu$ we assumed a fixed $\Gamma_{\rm av}$=4.5 and vary $\theta$ from 60$^{\circ}$ to 5$^{\circ}$ (green dots). The blue dots are the evolution of P$_{\rm jet}$ versus $\phi$, assuming a constant $\theta$=15$^{\circ}$ and varying $\Gamma_{\rm av}$ in the range 1--10. The green and blue dots show the increase of $\theta$ and $\Gamma_{\rm av}$, respectively. The red dots are the $\theta$ and $\Gamma_{\rm av}$ for which we can obtain a collimated jet ($\phi\leq10^{\circ}$) with a jet power P$_{\rm jet}\leq$0.05\,L$_{\rm Edd}$. }
    \label{fig:Pjet_phi}
\end{figure*}

As shown in Table \ref{tab:sed}, we obtained a good fit and jet and accretion flow parameters. However the {\sc ishem} model with an electron distribution index  p=2.5 predicts a flux in the hard X-rays which is about one order of magnitude below the data and cannot reproduce the soft $\gamma$-ray excess detected with SPI. However, as shown in the upper panel of Figure ~\ref{fig:sed} (blue dotted line), assuming a flatter electron energy distribution ($p$=2.1), makes the optically thin synchrotron emission harder, so matching the observed data.\\
Possible combinations of reasonable jet parameters corresponding to the best-fit model are shown in Table \ref{tab:sed}. Of course, as mentioned above, the  parameter degeneracy of {\sc ishem} implies that many different combinations of parameters can fit the data equally well. 
For this reason the individual statistical uncertainty on each of these parameters is not meaningful and was not calculated. The degeneracy is illustrated  in Figure \ref{fig:Pjet_phi}. Each point in this figure represents, a combination of jet parameters that produces the best-fit of the jet SED assuming $p=2.1$.  The lines show how the jet power (P$_{\rm jet}$) and the jet half-opening angle ($\phi$) change to keep a constant SED when the jet bulk Lorentz factor $\Gamma_{\rm av}$ or the inclination  $\vartheta$ are varied (while all the other parameters are kept constant).  

In the case of the $p=2.1$ model the {\sc xspec} fit provided only an upper limit on the frequency shift $\Delta\nu$ so the issue of degeneracy is even more stringent and we also have to consider combinations of parameters that are lower than this limit as they produce statistically comparable fits.  
We find that for a given $\vartheta$ and $\Gamma_{\rm av}$, a smaller shift $\Delta\nu$ results in an increasing of both $\phi$ and \pj~(black bullet in Fig. \ref{fig:Pjet_phi}).
Figure \ref{fig:Pjet_phi} shows that, for a fixed $\Delta\nu$, small jet-opening angles (low $\phi$) require large jet Lorentz factors (higher than a few). In this regime the jet power is rather insensitive to jet Lorentz factor and opening angle. A decrease of the mean Lorentz factor requires a larger jet-opening angle (high $\phi$) with a slightly higher jet power (blue dots in Fig. \ref{fig:Pjet_phi}).  When  $\Gamma_{\rm av}$ decreases below  $4.5$ both the jet power \pj~and opening angle $\phi$ start to increase significantly. When the jet opening angle exceeds $\sim$ 70$^{\circ}$, the jet power must increase substantially to maintain the correct flux. When $\Gamma_{\rm av}\sim$2, the jet opening angle reaches an almost constant value $\sim$100$^{\circ}$ and \pj~increases steeply.
The decreasing $\vartheta$ at constant $\Gamma_{\rm av}$, implies slightly larger opening-angles and  significantly lower jet power (green dots in Fig. \ref{fig:Pjet_phi}). Reasonably low power requires a small inclination.\\ 
Even though several parameter combinations can reproduce our data, these are not necessarily physically acceptable. 
Because compact jets are expected to be collimated, we favour solutions involving  opening angles lower than 10$^{\circ}$ \citep{miller-jones06}.
The model of \cite{kording06} provides a simple estimate of the mean value of the jet power using the unabsorbed X-ray luminosity \citep[see ][]{drappeau15,peault19}:
\begin{equation}\label{eq:power_jet}
    P_{jet}\approx 0.436\left[\frac{L_{[2-10\,keV]}}{L_{\rm Edd}}\right]^{\frac{1}{2}}L_{\rm Edd}.
\end{equation}
The unabsorbed X-ray luminosity L$_{[2-10\,keV]}\sim$0.01\,L$_{\rm Edd}$, resulting in a jet power of about 0.05\,L$_{\rm Edd}$. It is worth noticing that equation \ref{eq:power_jet} is based on a number of assumptions (e.g. a not efficient accretion flow, the jet power corresponds to a constant fraction of the accretion rate, the variation of the accretion rate across the state transition is smooth and relatively slow) which are not necessarily true for \grs. 
Therefore, the estimate gives only the approximate order of magnitude for the jet power. \\
These constraints in turn favour jets with $\Gamma_{\rm av}\geq$4.5 and low jet inclination $\vartheta\leq$20$^{\circ}$ (red dots in figure \ref{fig:Pjet_phi}).  

\section{The cooling break issue}

During the gas expansion in the jet, the electron distribution is induced to evolve under the influence of the acceleration and the cooling of the electrons. The electrons are affected by two types of cooling: the adiabatic cooling caused by the expansion of the jet in the interstellar medium, and the radiative cooling caused by the emission of synchrotron photons.
These two cooling mechanisms induce  two different slopes for the electron energy distribution. At low electron energies adiabatic expansion dominates while at higher energies radiation losses constitute the main cooling mechanisms.  This leads to a steepening  of  the lepton energy distribution at an energy corresponding to the transition between these two cooling regimes. This causes the appearance of a spectral  break in the  optically thin synchrotron spectrum.  The presence of such  a cooling break may strongly affect the high energy emission of the jet.
Up to now, a direct detection  of the cooling break has only been obtained for \maxi~\citep{russell14}, since it is usually hidden beneath other components. \\
The current version of the {\sc ishem} model does not take into account the radiation losses and therefore does not account for the presence of the cooling break. 
Nevertheless we can obtain a simple estimate of the energy of the cooling break in order to check whether it would affect our fit of the SED of \grs.
Assuming a constant injection of electrons described by a power law and constant adiabatic cooling during ejection, the stationary electron distribution as a function of energy $\gamma$, can be written as
\begin{equation}\label{eq_distr}
    N(\gamma)=\frac{S_{0}\tau_{a}}{p-1}\frac{ \gamma^{-p}}{\frac{\tau_{a} \gamma}{\tau_{s}} +1},
\end{equation}
where $\tau_{a}$ and $\tau_{s}$ are the characteristic adiabatic and synchrotron cooling times, respectively, defined according to the following relationships \citep{rybiky}:
\begin{equation}
    \tau_{a}=\frac{3z}{2 \Gamma_{av} \beta_{av} c}
\end{equation}
\begin{equation}
    \tau_{s}=\frac{6\pi m_{e}c}{\sigma_{\rm T}B^{2}_{j}}
\end{equation}
where $z$ is the position of the shock along the jet, $\beta_{av}=\sqrt{1-\frac{1}{\Gamma^{2}_{av}}}$ and $B_{j}$ is the magnetic field in the jet.
Depending on the values  chosen for $\tau_{a}$ and $\tau_{s}$, different behaviours are observed.
From equation \ref{eq_distr} we see that the electron energy at which the slope of the electron energy distribution changes is around:  
\begin{equation}\label{eq:cooling_break}
    \gamma_{b}(t)=\frac{\tau_{s}}{\tau_{a}}
\end{equation}
The {\sc ishem} simulations provide the magnetic field profile along the jet and this allows us to estimate at the typical distance z at which most of the high energy synchrotron emission is produced  and then estimate the magnetic field in this region.  From this we can estimate $\gamma_{b}$.  The electrons at this energy produce synchrotron photons at a typical frequency: 
\begin{equation}
    \nu_{c}=\frac{3eB_{j}}{4\pi m_{e}c}\gamma_{b}^{2}
\end{equation}
which should correspond approximately to the observed frequency of the cooling break. 
In the previous fitting procedure, we have assumed that the slope of the optically thin synchrotron emission is constant up to $\sim10$\,MeV, which implies that the cooling break should be at  energies higher than 10 MeV. However, from the above estimates and for the best-fit parameters given in Table \ref{tab:sed} this is expected around $10$\,keV ($\sim10^{18}$\,Hz).\\
In order to evaluate the effects of the cooling break in our fit, we have added a break in the {\sc xspec} local jet {\it ish} model. Figure \ref{fig:sed} (upper panel, black dash-dotted line) shows that the optically thin synchrotron emission softens above the break at 10\,keV. Therefore, the   
 {\sc ishem} model with $p=2.1$ can no longer reproduce the soft $\gamma$-ray emission of \grs~once the presence of a cooling break is accounted for.
Nevertheless, it is  still possible to reproduce the emission above 200\,keV with jet synchrotron emission but this would requires a much harder index of the electron distribution ($p<2$). We fitted our data assuming a harder electron distribution ($p=1.8$) and a break at 10\,keV.
As shown in the bottom panel of Figure ~\ref{fig:sed} (blue dotted line), the optically thin synchrotron emission  matches  the observed data.
The  best-fit  parameters  and a possible combination of reasonable jet parameters corresponding to the best-fit model are  shown  in  Table  \ref{tab:sed2}.

\begin{table}
	\centering
	\caption{Best-fit values of the SED by using the irradiated disc ({\it diskir}) model plus the jet internal shock emission model ({\sc ish}) taking into account a cooling break ($\nu_{c}$) at 10\,keV. The parameters of the simulation were obtained for a black hole of 4.9\,M$_{\odot}$ at 2.4\,kpc and assuming a jet with an electron distribution $p=1.8$.}
	\label{tab:sed2}
	\begin{tabular}{ll}
		\hline
		$p$                            & 1.8 \\
		$\nu_{c}$ (keV)                & (10) \\
		$\Delta\nu$                    & ($10^{-1.5}$)\\
		\hline
		N$_{\rm H}$ ($10^{22}$ \cm)    & 0.64$^{+0.02}_{-0.02}$\\
		E(B-V)                         & 0.93$^{+0.28}_{-0.12}$\\
		kT\_disk ($10^{-2}$\,keV)      & 3.58$^{+1.52}_{-1.41}$\\
		$\Gamma$                       & 1.66$^{+0.01}_{-0.01}$\\
		\kte (keV)                     & 49$^{+2}_{-2}$\\
		L$_{c}$/L$_{d}$                & >4.14\\
		fout ($\times 10^{-3}$)        & <0.02\\
		logrout                        & <2.73\\
		$\chi$(dof)                    & 1.00(430)\\
				 &\\
		$\vartheta$                    & $6^{\circ}$\\
		$\Gamma_{av}$                  & 10\\
		$\phi$                         & 3.67\\
		\pj~ (L$_{\rm E}$)             & 0.05\\
		\hline
	\end{tabular}
\end{table}

\begin{figure*}
	\includegraphics[width=17cm]{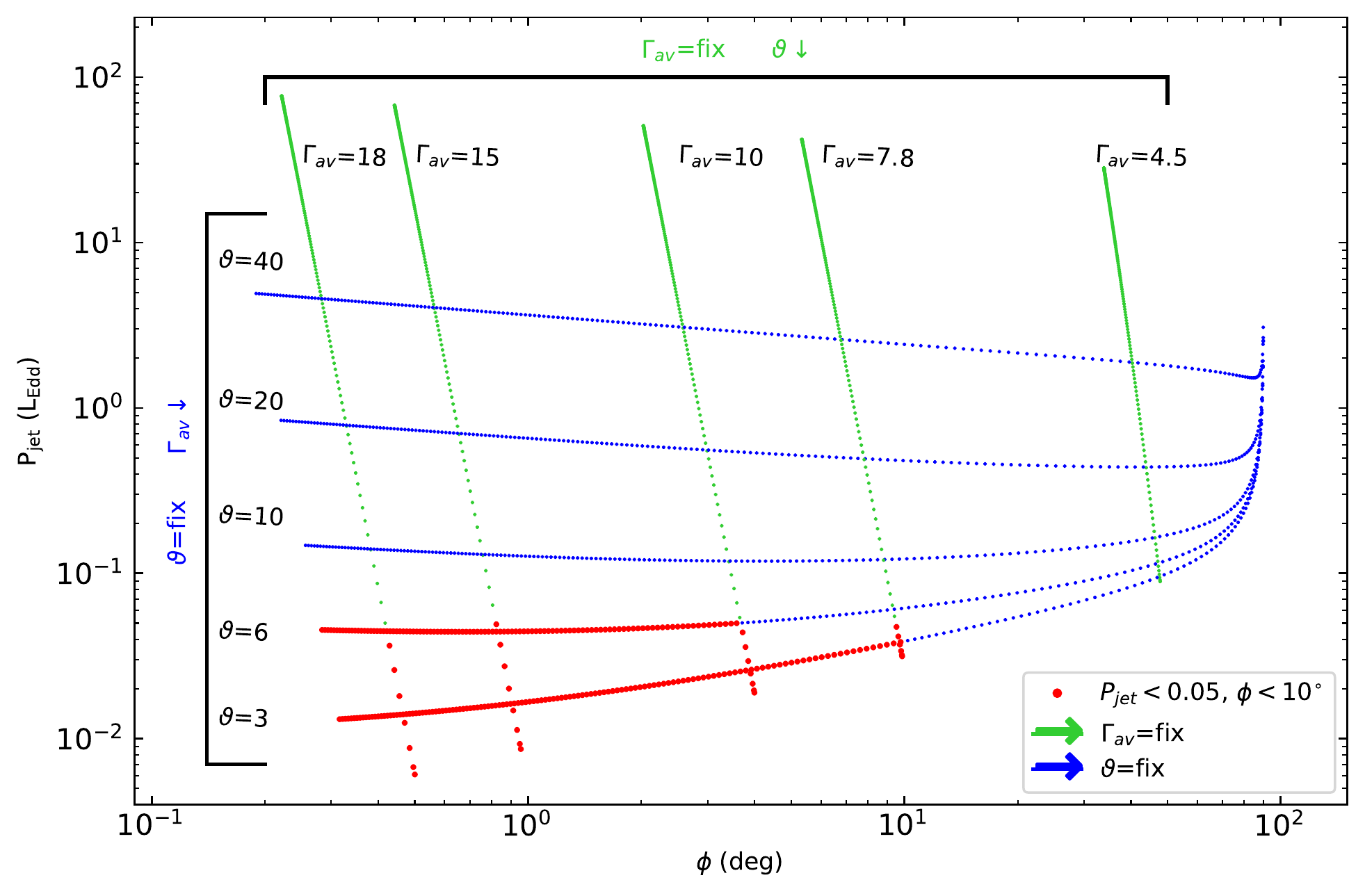}
    \caption{Evolution of the jet power (P$_{\rm jet}$) versus the jet opening angle ($\phi$) as a function of the mean Lorentz factor $\Gamma_{\rm av}$ and the jet inclination angle $\theta$ assuming $p=1.8$. We investigated the possible parameter combinations assuming a fixed $\Gamma_{\rm av}$=4.5, 7.8, 10, 15 and 18 and a variable $\theta$ (green dots). The blue dots are the evolution of P$_{\rm jet}$ versus $\phi$, assuming a constant value of $\theta$ (3$^{\circ}$, 6$^{\circ}$, 10$^{\circ}$, 20$^{\circ}$ and 40$^{\circ}$) and varying $\Gamma_{\rm av}$ in the range 1--20. The red dots are the $\theta$ and $\Gamma_{\rm av}$ for which we can obtain a collimated jet ($\phi\leq10^{\circ}$) with a jet power P$_{\rm jet}\leq$0.05\,L$_{\rm Edd}$. }
    \label{fig:Pjet_phi_1.8}
\end{figure*}

In Figure \ref{fig:Pjet_phi_1.8} we show the combination of jet parameters that produces the best-fit of the jet SED assuming $p=1.8$ and a break at 10\,keV. As in Figure \ref{fig:Pjet_phi}, also in this case the lines show how the jet power (P$_{\rm jet}$) and the jet half-opening angle ($\phi$) change when the jet bulk Lorentz factor $\Gamma_{\rm av}$ or the jet inclination $\vartheta$ vary. On the other hand, in this figure the shift frequency parameter has been kept frozen ($\Delta\nu=10^{-1.5}$).
In particular, the green dots in Figure \ref{fig:Pjet_phi_1.8} show how P$_{\rm jet}$ and $\phi$ change assuming constant values of $\Gamma_{\rm av}$, namely 4.5, 7.8, 10, 15 and 18, when  $\vartheta$ varies. Then, we fixed $\vartheta$ to values of 3$^{\circ}$, 6$^{\circ}$, 10$^{\circ}$, 20$^{\circ}$ and 40$^{\circ}$, allowing $\Gamma_{\rm av}$ to vary (blue dots in Fig. \ref{fig:Pjet_phi_1.8}).\\
Similarly to what we have obtained assuming $p=2.1$ (see Section \ref{jet_parameter} and Fig. \ref{fig:Pjet_phi}),  a decrease of the mean Lorentz factor requires larger $\phi$ values with a slightly higher \pj~ (blue dots in Fig. \ref{fig:Pjet_phi_1.8}). Then, decreasing $\vartheta$ at constant $\Gamma_{\rm av}$, implies slightly larger $\phi$ and  significantly lower \pj~ (green dots in Fig. \ref{fig:Pjet_phi_1.8}). Also in this case we favour solutions involving $\phi<10^{\circ}$ and \pj$\leq 0.05\,L_{\rm Edd}$ (red dots in Fig. \ref{fig:Pjet_phi_1.8}). However this region of the parameter space also implies  extreme jet Lorentz factors $\Gamma_{\rm av}\geq$10 and very small inclination  $\vartheta\leq$6$^{\circ}$ which we believe is unlikely. Overall, once the cooling break is included in the model we are not able to reproduce the data with 'reasonable' parameters.\\

\section{Discussion}

The simultaneous multi-wavelength campaign performed on \grs~allowed us to explore the possible origin of its soft $\gamma$-ray emission. 
First, we fitted the X/$\gamma$-ray spectrum with both the unmagnetised ({\sc eqpair}) and the magnetised ({\sc belm}) hybrid Comptonisation models.
The fits showed that the non-thermal Comptonizing electrons can be responsible for the excess above 200\,keV.
The best-fit with {\sc belm} in the limit of pure SSC, allowed us to estimate an upper limit on the magnetic field of the corona of about 1.5$\times$10$^{6}\left(\frac{20\,R_{\rm g}}{R}\right)$\,G. 
Furthermore, we found that the energy density  of the magnetic field is not sufficient to explain the radiation observed. In fact, the ratio U$_{B}$/U$_{R}$ (in terms of compactness) shows that the system is in a condition of sub-equipartition. A similar result was also derived for \Cyg~in the HS \citep{malzac12,delsanto13}.

The results from \cite{sobolewska11} indicate  that for luminosities higher than $\sim$1\%\,L$_{\rm Edd}$ the disc should supply the main seed photon contribution, while at lower luminosity synchrotron seed photons should dominate. The transition between these two regimes of Comptonisation occurs around a luminosity that is comparable to the  estimated bolometric luminosity of \grs~(L$\sim$0.01\,L$_{\rm Edd}$).
Indeed, we observed that the  Comptonized spectrum of \grs~can be explained  without soft seed photons from the accretion disc. 
If there would be an additional contribution of the seed photons from the disc, the magnetic field value should be significantly lower than the upper limit of $1.5 \times 10^6\left(\frac{20\,R_{\rm g}}{R}\right)$\,G we obtained in the hypothesis of pure synchrotron seed photons.\\

We also investigated whether this soft $\gamma$-ray emission could be due to the jet, thus we fitted the  SED of \grs~(from radio to $\gamma$-ray) with the the internal shock emission model {\sc ishem}.
We found that the jet synchrotron emission could explain the soft $\gamma$-ray emission if the index of the electron distribution is flat enough ($p\simeq 2.1$).
The results support a jet with an average bulk Lorentz factor ($\Gamma_{\rm av}$) higher than 4.5 and a maximum jet inclination angle of 20$^{\circ}$ assuming reasonable values of jet opening angle \citep[$\phi\leq$10$^{\circ}$, see ][]{miller-jones06}. 
The $\Gamma_{\rm av}$ we have found for \grs~is higher than the upper  limit of 2 derived from the L$_{X}\propto$L$_{R}^{0.7}$ correlation \citep{gallo03}. Even though the constraints on the bulk Lorentz factors of compact jets are few, it was shown that this correlation does not exclude high bulk Lorentz factors for BHBs jets \citep{heinz04}. Assuming that the low observed opening angles are a consequence of the transverse Doppler effect, and not due to some form of external confinement (e.g. magnetic or pressure confinement),  \cite{miller-jones06} showed that XRBs can produce relativistic jets with bulk speeds up to $\Gamma_{\rm av}\sim 10$, similarly to what is found in AGNs.
In any case, direct constraints on this parameter have been reported in only a few cases. From the IR/X-ray lags \cite{casella10} inferred a lower limit $\Gamma_{\rm av}\geq2$ for \gx~, while \cite{tetarenko19} constrained $\Gamma_{\rm av}\sim2.6$ for \Cyg~through a radio timing analysis. Additional constraints have been reported recently by \cite{saikia19}. These authors derived for a sample of nine BHBs  bulk Lorentz factors lying in the range 1.3-3.5.
Using the {\sc ishem} model, \citet{peault19} inferred that the jet Lorentz factor of the compact jet in \maxi~decreases  with the hardness of the X-ray spectrum and reaches values as high as $\Gamma_{av}\sim$16 in intermediate states. \\

However this jet  model does not self-consistently take into account  the cooling break in the jet electron distribution  that we expect should be present in this source and induce a spectral steepening  around 10\,keV. This in fact requires a harder index of the electron distribution ($p\leq2$). Indeed, we find that assuming $p=1.8$ and a cooling break at 10\,keV it is still possible to reproduce the soft $\gamma$-ray excess, although quite extreme jet parameters ($\Gamma_{av}\geq 10$ and $\vartheta\leq 6^{\circ}$) are required. Furthermore,  such a hard electron distribution is difficult to reconcile with the standard shock acceleration mechanisms.
Similar results were obtained by \cite{zdz14} who modelled the $\gamma$-ray emission of \Cyg~using a different jet model \citep{zdz14m}. 
They obtained a hard electron index ($p\sim1.5$) which allows to the formation of a hard synchrotron spectrum able to model the MeV tail \citep{zdz14}, difficult to reconcile with the shock acceleration mechanisms.
Thus, the non-thermal Comptonisation process within the corona seems more likely than the jet synchrotron for the nature of the soft $\gamma$-ray emission in \grs, unless alternative acceleration mechanisms are invoked.

\section{Conclusions}
In this paper we have presented the analysis of the X/$\gamma$-ray broad band spectra and the Spectral Energy Distribution of the BHT \grs~during its bright hard state that occurred in 2017 February--March.  Similarly to what has previously been observed in several sources, we detected the presence of a spectral component above 200\,keV which is in excess of the thermal Comptonisation spectrum.\\
Our results can be summarised as follow:
\begin{itemize}
    \item the hybrid thermal/non-thermal Comptonisation models provide a good description of the hard X-ray spectrum of \grs. In particular, the emission above 200\,keV can be explained as due to Comptonisation from non-thermal electrons;
    \item applying the {\sc ishem} model to the radio and soft $\gamma$-ray data, we found that the emission above 200\,keV can be due to the jet when the electron energy distribution has a flat slope ($p=2.1$), which is consistent with shock acceleration mechanisms;
    \item however,  the spectral shape of the jet emission at high energy may be strongly affected by radiative cooling which is not included in the current version of {\sc ishem}.  We have estimated that in \grs, this would lead to the formation of a spectral cooling break  at roughly 10\,keV. Once such a break is accounted for, it is not  possible to reproduce the soft $\gamma$-ray emission with the jet model, unless the index of the electron energy distribution would be lower than 2 which is difficult to reconcile with the shock acceleration mechanisms. In addition this requires quite extreme  bulk Lorentz factor (higher than 10) and very small jet inclination angle (lower than $6^{\circ}$). 
\end{itemize}
Measurements of $\gamma$-ray polarization in BHBs would provide strong constraints on the emission processes and geometry of the flow causing the soft $\gamma$-ray emission observed in the hard state.

\section*{Acknowledgements}
We acknowledge financial contribution from the agreement ASI-INAF n.2017-14-H.0 and INAF main-stream (P.I. Belloni).
This work also received financial support from PNHE in France and  from the OCEVU Labex (ANR-11-LABX-0060) and the A*MIDEX project (ANR-11-IDEX-0001-02) funded by the `Investissement d'Avenir'  French government program managed by the ANR. JCAM-J is the recipient of an Australian Research Council Future Fellowship (FT140101082), funded by the Australian government.




\bibliographystyle{mnras}
\bibliography{biblio} 







\bsp	
\label{lastpage}
\end{document}